\documentclass[1p,preprint,10pt]{elsarticle}

\usepackage{amsmath}
\usepackage{amsfonts}
\usepackage{array}
\usepackage{xcolor}
\usepackage{xspace}
\usepackage{hyperref}
\usepackage{lineno}
\usepackage{listings}
\usepackage{tabularx}
\usepackage{tabulary}
\usepackage{makecell}
\usepackage{placeins}
\usepackage{graphicx}
\usepackage{amssymb}
\usepackage{todonotes}
\definecolor{linkcolor}{HTML}{399B03}
\definecolor{urlcolor}{HTML}{399B03}
\hypersetup{pdfstartview=FitH, linkcolor=linkcolor,urlcolor=urlcolor,colorlinks=true}
\hypersetup{
    unicode=false,          
    pdftoolbar=true,        
    pdfmenubar=true,        
    pdffitwindow=false,     
    pdfstartview={FitH},    
    pdfauthor={},         
    colorlinks=true,        
    linkcolor=blue,         
    citecolor=red,          
}
\usepackage{bm}

\newcommand{\Green}{\textnormal{\texttt{Green}}\xspace}
\newcommand{\WeakCoupling}{\textnormal{\texttt{WeakCoupling}}\xspace}

\newcommand{\lati}{\bm{i}}
\newcommand{\latj}{\bm{j}}
\newcommand{\latk}{\bm{k}}
\newcommand{\latl}{\bm{l}}

\newcommand{\tk}{\mathbf{k}}
\newcommand{\tr}{\mathbf{r}}
\newcommand{\tR}{\mathbf{R}}
\newcommand{\tq}{\mathbf{q}}

\newcommand{\V}{\tilde{V}}

\definecolor{darkblue}{rgb}{0,0,.6}
\definecolor{darkred}{rgb}{.6,0,0}
\definecolor{darkgreen}{rgb}{0,.6,0}
\definecolor{red}{rgb}{.98,0,0}

\definecolor{cblue}{HTML}{034694}

\lstnewenvironment{pylisting}[1][]
{\lstset{language=Python,numbers=none,#1}}{}

\journal{Computer Physics Communications}

\begin{document}

\begin{frontmatter}

\title{\Green/\WeakCoupling: Implementation of fully self-consistent finite-temperature many-body perturbation theory for molecules and solids}

\author[Michigan]{Sergei Iskakov\corref{author}}
\ead{siskakov@umich.edu}
\author[NewYork]{Chia-Nan Yeh}
\ead{cyeh@flatironinstitute.org}
\author[MichiganChem]{Pavel Pokhilko}
\ead{pokhilko@umich.edu}
\author[Michigan]{Yang Yu}
\ead{umyangyu@umich.edu}
\author[Michigan]{Lei Zhang}
\ead{lzphy@umich.edu}
\author[MichiganChem]{Gaurav Harsha}
\ead{gharsha@umich.edu}
\author[MichiganChem]{Vibin Abraham}
\ead{avibin@umich.edu}
\author[MichiganChem]{Ming Wen}
\ead{wenm@umich.edu}
\author[MichiganChem]{Munkhorgil Wang}
\ead{munkhw@umich.edu}
\author[MichiganChem]{Jacob Adamski}
\ead{adamskij@umich.edu}
\author[WestChester]{Tianran Chen}
\ead{tchen@wcupa.edu}
\author[Michigan]{Emanuel Gull}
\ead{egull@umich.edu}
\author[Michigan,MichiganChem]{Dominika Zgid}
\ead{zgid@umich.edu}

\cortext[author] {Corresponding author. Current address: Department of Physics,
University of Michigan, Ann Arbor, Michigan 48109, USA}

\address[Michigan]{Department of Physics, University of Michigan, Ann Arbor, Michigan 48109, USA}
\address[MichiganChem]{Department of Chemistry, University of Michigan, Ann Arbor, Michigan 48109, USA}
\address[NewYork]{Center for Computational Quantum Physics, Flatiron Institute, 162 5th Avenue, New York, NY 10010, USA}
\address[WestChester]{Department of Physics and Engineering, West Chester University, West Chester, PA 19383, USA}

\begin{abstract}
The accurate ab-initio simulation of molecules and periodic solids with diagrammatic perturbation theory is an important task in quantum chemistry, condensed matter physics, and materials science.
In this article, we present the \WeakCoupling module of the open-source software package \Green, which 
implements fully self-consistent diagrammatic weak coupling simulations, capable of dealing with real materials in the 
finite-temperature formalism. The code is licensed under the permissive MIT license.
We provide self-consistent GW (scGW) and self-consistent second-order Green's function perturbation theory (GF2) solvers, analysis tools, 
and post-processing methods.
This paper summarises the theoretical methods implemented and provides background, tutorials and practical instructions for running simulations.
\end{abstract}

\end{frontmatter}

\noindent {\bf PROGRAM SUMMARY}

\begin{small}
\noindent
{\em Program Title:} \Green/\WeakCoupling \\
{\em Developer's repository link:} \url{https://github.com/Green-Phys/green-mbpt} \\
{\em Programming language:} \verb*#C++#, \verb|CUDA|, \verb*#Python#\\
{\em Licensing provisions:} MIT License 
{\em External routines/libraries:}
\verb#MPI >= 3.1#, \verb#BLAS#, \verb#Eigen >= 3.4.0#, \verb#cmake >= 3.18#, \verb|cuBLAS|.\\
{\em Nature of problem:}
The simulation of periodic solids and molecules using diagrammatic perturbation theory\\
{\em Solution method:}
We present an open-source implementation of the fully self-consistent finite-temperature  many-body perturbation
theory formalism at the GW and second-order perturbation theory level.
\end{small}

\section{Introduction}\label{sec:introduction}

The accurate, ab-initio, and systematically improvable numerical simulation of realistic interacting quantum systems using methods beyond density functional theory (DFT)~\cite{Hohenberg64,Kohn65} is an active, challenging area of research in condensed matter physics, quantum chemistry, and materials science. These simulations offer a route to material-specific, predictive modeling of experimentally complex systems, where an understanding of both physics and chemistry is necessary for gaining fundamental insights and optimizing technological advances.

Diagrammatic perturbation theory~\cite{abrikosov2012methods,fetter2003quantum,Mahan00,MartinInteracting16,vanneck} provides a rigorous theoretical framework for these simulations. The central object of this theory, the one-body Green's function, allows for the evaluation of multiple important quantities, such as total energies~\cite{PhysRevB.62.4858}, one-body observables,  and spectral information such as band structures and band gaps that relate the simulation results to experiments such as angle resolved photoelectron spectroscopy (ARPES)~\cite{Arpes_review}.
The evaluation of two-body observables, other than the total energy, requires either solving the Bethe-Salpeter equation~\cite{PhysRev.84.1232,PhysRevB.43.8044,Bickers2004} to obtain the two particle Green's function or applying the thermodynamic Hellmann–Feynman theorem~\cite{Pokhilko:local_correlators:2021}.
The computational access to experimentally relevant observables makes Green's function methods valuable for a broad range of applications.

Self-consistent diagrammatic theories, in particular $\Phi$- \cite{Luttinger60} or $\Psi$- \cite{Almbladh99} derivable methods, are particularly useful in this context since they guarantee thermodynamic consistency, conservation laws \cite{Baym61}, and -- in the absence of multiple stationary points in the $\Phi$- or $\Psi$- functional -- starting point independence achieved at self-consistency~\cite{Behnam_Farid_convergence_LW}.

Traditionally the domain of applicability of Green's function methods was limited to  weak interactions or high temperatures, 
where low-order perturbative  diagrammatic  expansions of the self-energy were shown to provide accurate results. Green's function methods, such as G$_0$W$_0$, when applied to semiconductors yielded  band gap values in much better agreement with experiment than the local density approximation (LDA), resulting in popularization of various approximations to the GW method in materials science~\cite{Schilfgaarde_Kotani_Faleev_GW_vs_LDA,Hybertsen_Louie_GW_1986}.

In recent years, the domain of applicability of Green's functions was extended to include moderately and strongly correlated systems, due to the introduction of Green's function embedding approaches~\cite{Georges96,Zgid17}, 
In these methods the low order perturbative expansion is performed for orbitals containing itinerant electrons, where correlations are weak, while non-perturbative corrections are introduced for a handful of orbitals containing localized electrons displaying strong correlations. 
Embedding methods such as the local density approximation with dynamical mean-field theory (LDA+DMFT)~\cite{Kotliar06} proved successful in qualitatively describing metal-to-insulator transition (MIT) in many strongly correlated perovskites.

The framework of many-body perturbation theory was developed already in the 1950s and 1960s \cite{Bohm53,Luttinger60,Hedin65} and its applications to  molecules~\cite{He_nat_orb_1969} and solids started in the late 1960s~\cite{AMO_physcis_GF}. However, the methods employed, such as GW,~\cite{Hedin65,Ferdi_1998_GW_review,reining_gw_2018,Rinke_GW_compendium} frequently contained severe approximations when applied to realistic systems or were applied without significant approximations but to only to model Hamiltonians.
Among the most common approximations employed were: performing only the first iteration of GW, thus truncating the bold diagrammatic series at G$_0$W$_0$; performing partial self-consistency schemes~\cite{vanschilfgaarde_quasiparticle_2006,Schilfgaarde_Kotani_Faleev_GW_vs_LDA,kutepov_electronic_2012,kutepov_linearized_2017,garcia-gonzalez_self-consistent_2001,PhysRevB.34.5390}, introducing a restriction of orbital space (e.g. by excluding most of the virtual orbitals~\cite{WEST_2015}); approximating the self-energy matrix by excluding off-diagonal elements (quasiparticle approximation)~\cite{Hedin65,Hedin_Lundqvist_1971}; restricting frequencies considered (by introducing plasmon-pole~\cite{Lundqvist1967a,Lundqvist1967b,Hedin_Lundqvist_1970,Overhauser_1971,Hybertsen_Louie_GW_1986,generalized_plasmon_pole_Farid,plasmon_pole_accuracy} or static approximations~\cite{static_dielectric_88,Bechstedt1992}).

The difficulty in executing the Green's function framework fully ab-initio is exemplified by the fact that while initial applications to model Hamiltonians describing systems as complicated as oxides started in the 1970s and were in the full swing delivering approximate descriptions of semiconductors in the 1980s~\cite{Hybertsen_Louie_GW_1986}, the fully self-consistent GW method without any approximations for electron gas was only executed in the 1997~\cite{scgw_electron_gas_1997}.

 This slow progress towards reaching fully ab-initio Green's function results for large realistic problems can be rationalized after considering the challenges present in realistic periodic systems such as (i) a wide Hamiltonian spectrum resulting in large frequency (time) grids when both core and virtual orbitals necessitate description, (ii) the difficulty of performing self-consistent iterations (resulting in an evaluation of bold diagrammatic series) (iii) the high computational scaling of weakly correlated methods such as GW and second order Green's function (GF2) scaling as $\mathcal{O}(n^6)$~\footnote{GW scaling of $\mathcal{O}(n^6)$ is present in the finite temperature Matsubara formalism when all the elements (diagonal and off-diagonal) of self-energy are evaluated and no decomposition of two-body integrals is employed. With the density fitted or Cholesky decomposed integrals, this scaling drops to $\mathcal{O}(n^4)$. A further reduction in scaling can be achieved when only diagonal elements of self-energy are evaluated.} and $\mathcal{O}(n^5)$, respectively, where $n$ is the number of orbitals in the problem, (iv) lack of numerical tools (up to 1990s) such as Cholesky decomposition~\cite{https://doi.org/10.1002/qua.560120408,10.1063/1.1578621}, density fitting~\cite{10.1063/1.1679012,10.1002/qua.560120408,doi:10.1021/ct200352g,PhysRevB.71.073103} , or tensor hypercontraction (THC)~\cite{Martinez:THC-MP2:2012} for two-body Coulomb integrals enabling a significant reduction of computational scaling  without resorting to uncontrolled diagrammatic series truncations or approximations. 

The fully ab-initio, rigorous evaluation of the many-body Green's function methods without any ad hoc approximations and with controllable accuracy for large realistic systems is only now becoming possible.
This is due to the substantial method development that happened within last 15 years ranging from adaptive grids \cite{Boehnke11,Kananenka16,Gull18,Shinaoka17,Kaye22}, decompositions of Coulomb integrals that proved to be accurate enough for Green's function methods while reducing the computational cost by orders of magnitude, convergence acceleration schemes~\cite{Pokhilko22,PhysRevB.108.155116}, development of full self-consistency in finite temperature Green's function approaches, and GPU acceleration that  was necessary to enable such simulations routinely and reliably.

Almost all practical diagrammatic  self-consistent methods are implemented in the finite-temperature Matsubara 
framework (on the imaginary axis) necessitating the use of analytic continuation methods to obtain spectral information such as band gap and band structure.
While traditional methods either smooth data (such as the Maximum Entropy 
method~\cite{Jarrell96}), or possibly violate causality (such as the Pad\'{e} method \cite{Vidberg77}), 
recent theoretical progress resulting in the Nevanlinna \cite{Fei21,Fei21A}, projection-estimation-semidefinite relaxation (PES)~\cite{Huang23} and 
Prony~\cite{YING2022111549,Ying2022,Zhang23} methods enables causal, accurate, and systematically improvable continuations. 

While many publicly available codes evaluating weakly correlated Green's functions such as GW are available, very few codes enable a fully ab-initio rigorous evaluating of Green's function containing the newest developments described above and providing a good control of the accuracy while being  available in readily accessible user-friendly simulation packages. 
\Green/\WeakCoupling aims to provide an implementation of these methodologies as a permissively licensed open-source software package, together with tutorials, documentation, and examples, to enable diagrammatic materials calculations with self-consistent techniques, along with additional post-processing tools.

\section{\label{sec:theory}Theory}

\Green/\WeakCoupling provides an approximate solution for realistic quantum systems of interacting electrons using Green's function based many-body perturbation theory methods.
The interacting quantum  system is described  by a second quantized Hamiltonian of the form  
\begin{align}
    H &= \sum_{ij,\sigma} H^{\tk_i\tk_j(0)}_{ij} c^\dagger_{\tk_ii\sigma}c_{\tk_jj\sigma} + \frac{1}{2 N_k} \sum_{ijkl,\sigma\sigma'} U^{\tk_i\tk_j\tk_k\tk_l}_{ijkl} 
    c^\dagger_{\tk_ii\sigma} c^\dagger_{\tk_kk\sigma'} c_{\tk_ll\sigma'}c_{\tk_jj\sigma}. \label{Eqn:Hamiltonian}
\end{align}
Here, the operators $c^\dagger$ ($c$) create (destroy) electrons with spin $\sigma$, momentum $\tk_i(\tk_j,\tk_k,\tk_l)$ and the atomic orbital index $\lati (\latj, \latk, \latl)$~\cite{szabo1996modern}. For molecular systems momentum index is omitted. Given a set of single-particle basis states $\psi^{\tk_i}_i(\tr)$, the one-body Hamiltonian $H^{\tk_i\tk_j(0)}_{ij}$, in the simplest case, is defined as 
\begin{align}
    H^{\tk_i\tk_j(0)}_{ij} = \int d\tr \psi^{\tk_i*}_i(\tr)  \left(-\frac{\hbar^2}{2m} \nabla^2_\tr  - V_{ext} (\tr)\right)  \psi^{\tk_j}_j(\tr),
\end{align}
and describes the motion of electrons in the external potential $V_{ext}(\tr)$.
The electron-electron repulsion\footnote{\Green/\WeakCoupling uses 
chemists notation (See Ref.~\cite{szabo1996modern}) for the electron-electron repulsion, as it's commonly used in many ab initio packages.} is
\begin{align}
U^{\tk_i\tk_j\tk_k\tk_l}_{ijkl} = \int d\tr_1d\tr_2 \psi^{\tk_i*}_i(\tr_1) \psi^{\tk_j}_j(\tr_1) \frac{1}{|\tr_1 - \tr_2 |} \psi^{\tk_k*}_k(\tr_2) \psi^{\tk_l}_l(\tr_2).
\label{eqn:Coulomb}
\end{align}
This rank-four Coulomb tensor is typically decomposed with the help of the density fitting~\cite{10.1063/1.1679012,10.1002/qua.560120408,doi:10.1021/ct200352g,PhysRevB.71.073103} or Cholesky decomposition~\cite{https://doi.org/10.1002/qua.560120408,10.1063/1.1578621} techniques into a product of two low-rank tensors as
\begin{align}
    U^{\tk_i\tk_j\tk_k\tk_l}_{ijkl} = \sum_{Q} \V_{\tk_ii,\tk_jj}(Q) \V_{\tk_kk,\tk_ll}(Q),
    \label{eqn:CoulombDF}
\end{align}
where $Q$ is an auxiliary index. The current version of \Green/\WeakCoupling uses PySCF~\cite{PYSCF} package to 
prepare one- and two-body integrals and we assume that single-particle particle basis functions 
$\psi^{\tk_i}_i(\tr)$ are either periodic Gaussian-type orbitals (for solids) or Gaussian-type 
orbitals (for molecules).

\Green/\WeakCoupling provides an approximate solution of system described by the Hamiltonian from Eq.~\ref{Eqn:Hamiltonian} within the Green's function language, where the main objects are imaginary-time or Matsubara-frequency-dependent electron Green's functions and electron self-energies expressed in terms of Feynman diagrams~\cite{mattuck1992guide}.

It is convenient to split the self-energy into static and time-dependent parts:
\begin{align}
\Sigma_{ij}(\tau) = \Sigma^\infty_{ij}\delta_{\tau} + \tilde{\Sigma}_{ij}(\tau),
\end{align}
where $\Sigma^\infty_{ij}$ corresponds to the static infinite-frequency limit of 
the self-energy, and the dynamical part $\tilde{\Sigma}_{ij}(\tau)$ encapsulates the dependence on imaginary time $\tau$, $0\leq\tau\leq\beta$, where $\beta=1/k_BT$ is the inverse of the physical temperature $T$.

The solution of the lowest-order approximation of diagrammatic perturbation theory, Hartree-Fock,  leads only to a static self-energy
\begin{align}
    \Sigma^{\infty,\tk}_{\sigma,ij} = \frac{1}{N_{k}}\sum_{\substack{\tk'\sigma'\\ab Q}}
    \V_{i\tk,j\tk}(Q)G^{\tk'}_{\sigma',ab}(0^-)\V_{b\tk',a\tk'}(Q) -
    \frac{1}{N_{k}}\sum_{\substack{\tk'Q\\ab}} \V_{i\tk,a\tk'}(Q) 
    G^{\tk'}_{\sigma,ab}(0^-)\V_{b\tk',j\tk}(Q),
    \label{eq:sigma_inf}
\end{align}
where $G^{\tk}_{\sigma,ij}(\tau)$ is the single-particle imaginary time Green's function.

Higher order approximations 
yield additional dynamical corrections. 
\Green/\WeakCoupling provides an implementation of two approximations beyond Hartree-Fock, namely the second-order self-consistent perturbation theory (GF2) and  self-consistent GW method (scGW).

\subsection{GF2 approximation}
GF2, also known as Second-order Born approximation~\cite{Born1926}, is a self-consistent conserving diagrammatic approximation that contains all the self-energy diagrams up to a second order in the interaction~\cite{Phillips14,Rusakov16,PhysRevB.100.085112}. Generally, it is considered accurate whenever gaps are large and interactions are weak but it is known to fail for metals. Unlike GW \cite{Hedin65} (cf. Sec.~\ref{sec:GW approximation}), GF2 is expanded in terms of bare Coulomb interactions, and it contains a second-order exchange term but does not include higher-order screening contributions.

The second-order self-energy in the imaginary time and momentum space with decomposed Coulomb interaction is 
\begin{align}\label{eq:sigma2}
\tilde{\Sigma}^{\tk,(2)}_{\sigma,ij}(\tau) &= -\frac{1}{N_{\tk}^3}\sum_{\substack{\tk_1\tk_2\tk_3\\klmnpq\\\sigma',QQ'}}
(\V^{\tk_1\tk}_{qj}(Q') \V^{\tk_2\tk_3}_{ln}(Q') - 
\V^{\tk_2\tk}_{lj}(Q')\V^{\tk_1\tk_3}_{qn}(Q')\delta_{\sigma\sigma'}) \\  \nonumber
\times \V^{\tk\tk_1}_{ip}(Q)& \V^{\tk_3\tk_2}_{mk}(Q)
G^{\tk_1}_{\sigma,pq}(\tau)G^{\tk_2}_{\sigma',kl}(\tau)
G^{\tk_3}_{\sigma',nm}(-\tau)\delta_{\tk+\tk_3,\tk_1+\tk_2},
\end{align}
with imaginary time Green's function $G^{\tk}_{\sigma,ij}(\tau)$. 

A detailed discussion on the GF2 approximation implemented in \Green/\WeakCoupling  is presented in Ref.~\cite{Rusakov16,PhysRevB.100.085112}.

\subsection{GW approximation}\label{sec:GW approximation}
The GW method sums an infinite series of so-called random phase approximation (RPA) or ``bubble" diagrams, thereby including screening
effects but neglecting the second-order exchange diagrams~\cite{Hedin65}.
Note that there are several variants of the GW approximation that correspond to different equations and additional approximations, including non-selfconsistent~\cite{PhysRevB.34.5390}, partially self-consistent~\cite{PhysRevB.54.8411}, quasiparticle approximated and quasiparticle self-consistent variants~\cite{PhysRevLett.99.246403}. \Green/\WeakCoupling implements fully self-consistent GW approximation~\cite{Yeh22,KUTEPOV2020107502}.

In scGW, the contribution to the dynamical part of the self-energy is
\begin{align}
    \tilde{\Sigma}^{(GW),\mathbf{k}}_{i\sigma,j\sigma}(\tau) = -\frac{1}{N_{k}}\sum_{\mathbf{q}}\sum_{ab} G^{\mathbf{k-q}}_{\sigma,ab}(\tau)\tilde{W}^{\mathbf{k},\mathbf{k-q},\mathbf{k-q},\mathbf{k}}_{ i a  b j}(\tau),
    \label{eq:sigma_gw}
\end{align}
with dynamical screened interaction expressed in the frequency space as
\begin{align}
&\tilde{W}^{\bold{k},\bold{k-q},\bold{k-q},\bold{k}}_{ijkl}(i\Omega_{n}) = 
\sum_{QQ'}\V^{\bold{k},\bold{k-q}}_{ij}(Q)\tilde{P}^{\bold{q}}_{QQ'}(i\Omega_{n})\V^{\bold{k-q},\bold{k}}_{kl}(Q')
\label{Eq:tilde_W_DF2}
\end{align}
where $\Omega_{n} = 2n\pi/\beta$ ($n \in \mathbb{Z}$) are bosonic Matsuabara frequencies. 
$P^{\tq}_{QQ'}(\Omega_n)$ is a renormalized auxiliary function
\begin{align}
\tilde{\bold{P}}^{\bold{q}}(i\Omega_{n}) = \sum_{m=1}^{\infty}[\tilde{\bold{P}}^{\bold{q}}_{0}(i\Omega_{n})]^{m} 
= [\bold{I} - \tilde{\bold{P}}^{\bold{q}}_{0}(i\Omega_{n})]^{-1}\tilde{\bold{P}}^{\bold{q}}_{0}(i\Omega_{n}),
\label{Eq:tilde_P_omega}
\end{align}
that is obtained from a bare auxiliary function
\begin{align}
\tilde{P}^{\bold{q}}_{0,QQ'}(i\Omega_{n}&) = \int_{0}^{\beta}d\tau \tilde{P}^{\bold{q}}_{0,QQ'}(\tau)e^{i\Omega_{n}\tau}, \label{Eq:P0t_to_P0w}\\
\tilde{P}^{\bold{q}}_{0,QQ'}(\tau) = -\frac{1}{N_{k}}\sum_{\substack{\tk \sigma \\abcd}}&\V^{\tk,\tk+\tq}_{da}(Q) G^{\tk}_{\sigma,cd}(-\tau)G^{\tk+\tq}_{\sigma,ab}(\tau)\V^{\tk+\tq,\tk}_{bc}(Q')\label{Eq:tilde_P0_tau}
\end{align}

For detailed derivation of the GW formalism implemented within \Green/\WeakCoupling with a non-relativistic Hamiltonian refer to Ref.~\cite{Yeh22}
A relativistic GW (X2C1e-scGW) within the exact two-component formalism (X2C1e)~\cite{Trond_X2C1e} is presented in Ref.~\cite{Yeh22B}.

\subsection{Self-consistency loop}\label{sec:sc}
\begin{figure}[tbh]
    \centering
    \includegraphics[width=.8\textwidth]{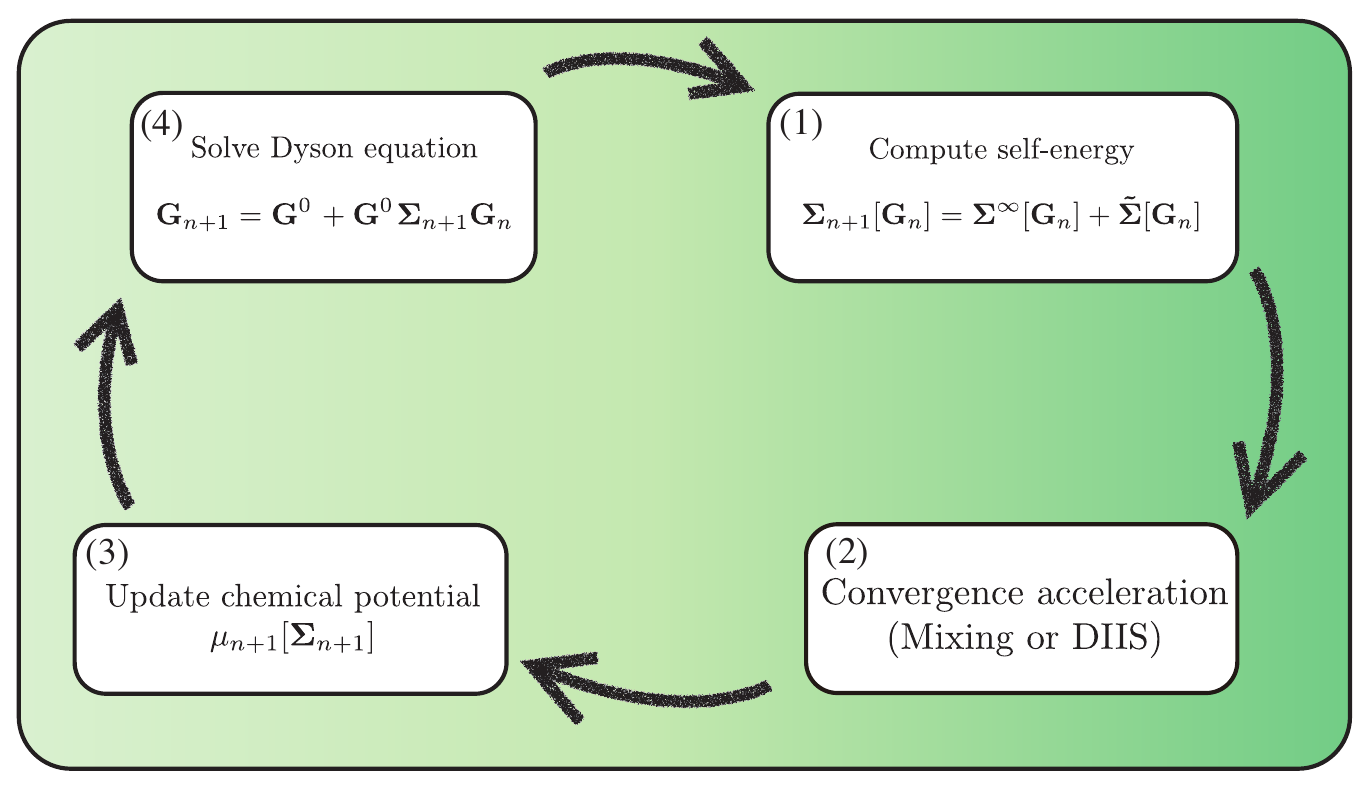}
    \caption{Schematic illustration of the self-consistency loop implemented in \Green/\WeakCoupling.}
    \label{fig:scloop}
\end{figure}

\Green/\WeakCoupling provides an implementation of the fully self-consistent many-body perturbation theory. 
Self-consistent iterations start from an initial guess for the Green's function $G$, usually either 
obtained from a ground-state mean-field solution or from the non-interacting limit, and proceed as following. (1) The static part (Eq.~\ref{eq:sigma_inf}) and the dynamic part (Eq.~\ref{eq:sigma2} or Eq.~\ref{eq:sigma_gw}) of the self-energy are evaluated. (2) To improve convergence, a convergence acceleration technique is be applied. (3) The chemical potential is adjusted to obtain a fixed number of particles. (4) A new approximation for the Green's function is calculated  via the Dyson equation and the simulation proceeds to step (1) until the desired accuracy of convergence has been reached. Fig.~\ref{fig:scloop} shows a schematic representation of the aforementioned procedure.

\subsection{Postprocessing}\label{sec:post}
\subsubsection{$\tk$-path interpolation}\label{sec:kinterp}
\Green/\WeakCoupling evaluates Green's functions and self-energies on a relatively small 
Monkhorst-Pack $\tk$-grid~\cite{PhysRevB.13.5188}. However, a visually appealing presentation of quantities 
such as the band-structure requires a dense $\tk$-mesh.

The frequency-momentum representation of the Green's function obtained by \Green/\WeakCoupling
is
\begin{align}
    G^{\tk}_{ij}(\omega_n) = \left[(i\omega_n + \mu)\mathbf{S}^\tk - \mathbf{H^{\tk,(0)}} - 
    \mathbf{\Sigma^{\infty,\tk}} - \mathbf{\tilde{\Sigma}}^{\tk}(\omega_n) \right]^{-1}_{ij},
    \label{eqn:dyson}
\end{align}
with Matsubara frequency $\omega_n = (2n+1)\pi/\beta$, chemical potential $\mu$, and 
overlap matrix $S_{ij}^\tk$.

To obtain the Green's function on a denser grid $\hat{\tk}$, we perform a simple $\tk$-space interpolation 
similar to a Wannier interpolation~\cite{RevModPhys.84.1419}, with the exception that we do not perform 
the transformation to a basis of maximally localized basis of Wannier functions.

In this approach, we first transform momentum dependent quantities $\hat{X}(\tk)$ into real space via Fourier transform,
\begin{align}
    X(\tR) = \frac{1}{N_{\tk}}\sum_{\tk} \hat{X}(\tk)e^{-i\tk\tR}.
\end{align}
Then, by applying the inverse transform, we project them back onto a finer $\hat{\tk}$-grid
\begin{align}
    \hat{X}(\hat{\tk}) = \sum_{\tR} X(\tR)e^{i\hat{\tk}\tR}.
\end{align}
If the real-space quantity $X(\tR)$ is well localized, which is usually the case when single-particle basis funcations $\psi^{\tk_i}_{i}(\textbf{r})$ are Gaussian-type orbitals, the resulting quantity will be approximated well. We found that the best results are obtained when only 
$\Sigma^{\infty,\tk}$ and $\tilde{\Sigma}^\tk$ are interpolated, and both the noninteracting Hamiltonian $H^0$ and 
the overlap matrix $S$ are directly evaluated on a finer $\hat{\tk}$-grid. Then the Green's function is evaluated using the Eq.~\ref{eqn:dyson} on a finer $\hat{\tk}$-grid. 

\subsubsection{Analytical continuation}

Standard finite-temperature perturbation theories are formulated on the imaginary axis allowing the 
direct access to static quantities such as the density matrix, free energy, or entropy. 
In contrast, quantities defined on real-frequency grids such as the spectral function require 
analytical continuation to the real axis of imaginary-time data~\cite{Jarrell96,Fuchs10} by solving the inverse problem of
\begin{align}
    G^{\sigma}_{ij}(\tau)= -\int d\omega \frac{A^{\sigma}_{ij}(\omega)e^{-\tau\omega}}{1+e^{-\beta\omega}}.\label{eqn:dos_tau}
\end{align}
The solution of the analytical continuation problem is ill conditioned and numerically unstable in 
the presence of noise~\cite{Jarrell96}. \Green/\WeakCoupling results are free of stochastic noise 
and methods suitable for noise-free data can be applied. 
\Green/\WeakCoupling provides an implementation of analytical continuation using the Nevanlinna 
analytical continuation method as described in Ref.~\cite{Fei21}. Several implementation of these method exists~\cite{10.21468/SciPostPhysCodeb.19,arxiv.2309.01407}.

\section{Installation}\label{sec:installation}
\subsection{Dependencies}
Green requires the following dependencies to be installed and available:
\begin{itemize}
    \item Eigen3 $>=$ 3.4.0. Eigen is a C++ template library for linear algebra: matrices, vectors, numerical solvers, and related algorithms \cite{Eigenv3}.
    \item MPI. Message Passing Interface (MPI) is a standardized and portable message-passing standard designed to function on parallel computing architectures \cite{mpi40}.
    \item HDF5 $>=$ 1.10.0. HDF5 is a high-performance data management and storage suite  \cite{HDF5}.
    \item BLAS. The BLAS (Basic Linear Algebra Subprograms) are routines that provide standard building blocks for performing basic vector and matrix operations \cite{BLAS}.
    \item CMake $>=$ 3.18. CMake is a tool to manage building of source code \cite{CMAKE}.
    \item CUDAToolkit $>=$ 11.1 (optional). CUDA is a parallel computing platform and programming model for general computing on graphical processing units (GPUs) \cite{CUDA08}.
    \item GMP (optional, for analytic continuation). The GNU Multiple Precision Arithmetic Library (GMP) is a library for arbitrary-precision arithmetic, operating on signed integers, rational numbers, and floating-point numbers \cite{GMP}
\end{itemize}

These packages are external to Green. Installation instructions are provided with these packages; many modern HPC environments provide some or all of these packages preinstalled for their users.

In addition, the following python dependencies are required for auxiliary scripts.
\begin{itemize}
    \item green-mbtools. The python tool suite for Green's-function-based many-body calculations using \Green Software Package.
    \item PySCF. The Python-based Simulations of Chemistry Framework (PySCF) is an open-source collection of electronic structure modules powered by Python \cite{PYSCF}.
    \item numba. Numba translates Python functions to optimized machine code at runtime using the industry-standard LLVM compiler library \cite{NUMBA}.
    \item spglib. SPGLib is a software library for crystal symmetry search \cite{Togo24}.
    \item ase. ASE is an Atomic Simulation Environment written in the Python programming language with the aim of setting up, steering, and analyzing atomistic simulations \cite{ASE}.
\end{itemize}
Python packages can typically be provisioned with the \lstinline[language=bash]{pip install} or \lstinline[language=bash]{pip3 install} command. \lstinline[language=bash]{pip install pyscf numba spglib ase green-mbtools} will install the required python modules on most platforms.

\subsection{Weak coupling}
The following instructions will download, build and install the CPU-only version of \Green/\WeakCoupling solver (replace \texttt{/path/to/install/directory} with the desired installation directory path; see below for the CPU/GPU version):
\begin{verbatim}
$ git clone https://github.com/Green-Phys/green-mbpt
$ mkdir green-mbpt-build
$ cd green-mbpt-build
$ cmake -DCMAKE_BUILD_TYPE=Release \
        -DCMAKE_INSTALL_PREFIX=/path/to/install/directory ../green-mbpt
$ make -j 4 && make test
$ make install
\end{verbatim}
This sequence of commands will create the installation directory and place the executable mbpt.exe under the bin directory of the installation path.

The following instructions will download, build and install the GPU version of the \Green/\WeakCoupling solver (replace \texttt{/path/to/install/directory} with the desired installation directory path):
\begin{verbatim}
$ git clone https://github.com/Green-Phys/green-mbpt
$ mkdir green-mbpt-build
$ cd green-mbpt-build
$ cmake -DCMAKE_BUILD_TYPE=Release                                \
   -DCUSTOM_KERNEL=GPU_KERNEL                                     \
   -DGREEN_KERNEL_URL="https://github.com/Green-Phys/green-gpu"   \
   -DGREEN_CUSTOM_KERNEL_LIB="GREEN::GPU"                         \
   -DGREEN_CUSTOM_KERNEL_ENUM=GPU                                 \
   -DGREEN_CUSTOM_KERNEL_HEADER="<green/gpu/gpu_factory.h>"       \
   -DCMAKE_INSTALL_PREFIX=/path/to/install/directory ../green-mbpt
$ make -j 4 && make test
$ make install
\end{verbatim}
The GPU version requires the CUDA toolkit \cite{CUDA08} to be available.

\subsection{Analytical continuation}

In addition to the \Green/\WeakCoupling dependencies, the analytical continuation code requires the GNU multiprecision arithmetics library \cite{GMP}.
The following instructions will download and build the analytical continuation package (replace /path/to/install/directory with the desired installation directory):
\begin{verbatim}
  $ git clone https://github.com/Green-Phys/green-ac.git
  $ mkdir build
  $ cd build
  $ cmake -DCMAKE_BUILD_TYPE=Release \
          -DCMAKE_INSTALL_PREFIX=/path/to/install/directory ../green-ac
  $ make -j 4 && make test && make install
\end{verbatim}

\section{Usage}\label{sec:usage}

\subsection{Input preparation}\label{subsec:input}
The first step of any simulation with \Green/\WeakCoupling is the preparation of the input data for the diagrammatic calculation. This entails the calculation of the  one-body integrals, the calculation of the (decomposed or density fitted) Coulomb integrals, and the determination of a starting density matrix. These components are generated from unit-cell geometry information, atom positions, the k-point discretization grid, as well as information about the basis.

We provide a convenient interface to the PySCF package \cite{PYSCF} to prepare this input by calling the python script \lstinline[language=bash]{python/init_data_df.py}, located in the installation path (for a simple example see section~\ref{sec:examples}). The following parameters are mandatory:

\begin{itemize}
    \item \lstinline[language=Python]!--a! -- path to a file containing the translation vectors of the primitive unit cell, in angstrom.
    \item \lstinline[language=Python]!--atom! -- path to a file containing the atom species and position of atoms in the unit cell in the standard xyz format, in angstrom.
    \item \lstinline[language=Python]!--nk! -- number of $\tk$-points in each direction. Assumed to be the same for all directions.
    \item \lstinline[language=Python]!--basis! -- Gaussian basis set description.
\end{itemize}

The \Green/\WeakCoupling interface to pySCF employs Gaussian-type orbitals as evaluated by PySCF \cite{PYSCF} as single-particle states. Two options for choosing a basis set are provided: a) using a built-in  
molecular~\cite{pyscf-mol-basis} or 
periodic~\cite{pyscf-pbc-basis} basis;
b) providing a file that contains desired basis set in the NWChem~\cite{NWCHEM} format. Users can either select a single basis set for all atoms in the calculation (\lstinline[language=Python]{--basis <basis set>}) or 
provide an individual basis for each atom species separately:

\lstinline[language=Python]{--basis <first atom> <first basis set> <second atom> <second basis>...}.

Basis sets with pseudopotential require the user to provide a pseudopotential for core electrons using the option \lstinline[language=Python]{--pseudo}. Users can either choose a built-in PySCF pseudopotential~\cite{pyscf-pp} or provide a file containing pseudopotentials in the NWChem format.

Users can use an external MolSSI BSE tool~\cite{MolSSI_BSE} for basis set conversions from other formats.

By default the script \lstinline[language=bash]{init_data_df.py} will generate a file 
\lstinline[language=bash]{input.h5} containing all necessary parameters and an initial 
mean-field solution of the system, as well as the one-body integrals.
In addition, the script will generate a directory named \lstinline[language=bash]{df_hf_int} which contains 
the decomposed Coulomb integrals of the system with integrable divergence at $(\tq + \mathbf{G})=0$ 
explicitly removed. If parameter \lstinline[language=bash]{--finite_size_kind} set to 
\lstinline[language=bash]{ewald} script will generate an additional directory named 
\lstinline[language=bash]{df_int} which contains the decomposed Coulomb integrals with integrable divergence
at $(\tq + \mathbf{G})=0$ replaced with an estimate using Madelung constant (see 
Sec.~\ref{subsubsec:finitesieze} for more details on finite-size corrections).

After the simulation is performed, some of the results, such as the single particle spectral function, are displayed along a high-symmetry path. To obtain a high-symmetry path, high-symmetry points on the path and the total number of points along the path are provided using the two parameters \lstinline[language=bash]{--high_symmetry_path} and \lstinline[language=bash]{--high_symmetry_path_points}. The option \lstinline[language=bash]{--print_high_symmetry_points} lists all available high-symmetry points for a system. The interpolation along the desired high-symmetry paths will perform a basic $\tk$-path interpolation (see Sec.~\ref{sec:kinterp}) for those points not on the momentum mesh of the calculation.

All \Green programs list their available options with the \lstinline[language=bash]{--help} option.

\subsection{Solution of the many-body perturbation theory equations}\label{subsec:solve}
The many-body perturbation theory equations are solved by the program \lstinline[language=bash]{mbpt.exe}  located in the \texttt{bin} directory of the installation path. It is advantageous to perform this simulation on a machine with multiple cores or nodes (using the CPU MPI implementation) or on GPUs (using the GPU implementation), rather than on a local workstation or laptop. Minimal parameters needed to run weak-coupling simulations are as follows:

\begin{itemize}
    \item \lstinline[language=bash]{--BETA} -- The inverse temperature, in inverse energy units
    \item \lstinline[language=bash]{--scf_type} -- The type of the diagrammatic approximation,  either
    \lstinline[language=bash]{HF} (for Hartree Fock), \lstinline[language=bash]{GF2} (for fully self-consistent second-order perturbation theory), or \lstinline[language=bash]{GW} (for fully self-consistent GW)
    \item \lstinline[language=bash]{--grid_file} -- the path to a file containing the non-uniform grid information. The program will automatically check three possible locations in the following order:
    \begin{itemize}
        \item current directory or absolute path
        \item \lstinline[language=bash]{<installation directory>/share}
        \item the build directory of mbpt code
    \end{itemize}
    As a default, we provide intermediate representation (IR \cite{Shinaoka17}, \lstinline[language=bash]{ir} subdirectory) and Chebyshev (\cite{Gull18}, \lstinline[language=bash]{cheb} subdirectory) grids for nonuniform imaginary time representation.
\end{itemize}

After successful completion, results will be written to a file located at \lstinline[language=bash]{--results_file} (by default set to \lstinline[language=bash]{sim.h5}) Additional parameters and their default values are listed by calling the porgram with the \lstinline[language=bash]{--help} option.

Provided the option \lstinline[language=bash]{--jobs WINTER} is set and the input data contains information about a desired high-symmetry path, the diagonal part of the Green’s function is evaluated on the path, and results are stored in the \lstinline[language=bash]{G_tau_hs} group in the file specified as \lstinline[language=bash]{--high_symmetry_output_file} (by default set to \lstinline[language=bash]{output_hs.h5}).

\subsubsection{Finite-size corrections}\label{subsubsec:finitesieze}
In periodic calculations the infinite crystal is approximated by a finite crystal with periodic boundary conditions defined on a discrete set of momentum points. This finite-size approximation introduces finite-size errors. \Green/\WeakCoupling provides the following options to control finite-size errors:

In Hartree-Fock calculations we compute explicit Ewald corrections using the Madelung constant following  Baldereschi's scheme~\cite{PhysRevB.34.4405}.

In GW calculations we provide two options:
\begin{itemize}
    \item \lstinline[language=bash]{EWALD_INT} enables explicit Ewald correction using integrals with divergent part of Coulomb kernel replaced with supercell Madelung constant~\cite{10.1063/1.1926272,PhysRevB.80.085114};
    \item \lstinline[language=bash]{EXTRAPOLATE} computes an extrapolation of the dielectric matrix to obtain the $\mathbf{G} = 0$ contribution. This requires the calculation of a smooth interpolant for the Coloumb integral~\cite{10.1063/1.1926272,PhysRevB.80.085114,Yeh22}.
\end{itemize}

In the GF2 calculations, the divergent part of the Coulomb kernel is replaced with the supercell Madelung constant via the Ewald correction~\cite{10.1063/1.1926272,PhysRevB.80.085114}. The finite-size corrections will be applied if the file 
\lstinline[language=bash]{df_ewald.h5} is present in the integrals directory. 

The finite-size corrections for the second-order and GW approximations 
require the generation of additional input data. In the initialization script, this is controlled by the 
\lstinline[language=bash]{--finite_size_kind} parameter which takes the following values
\begin{itemize}
    \item \lstinline[language=bash]!ewald! -- will generate the \lstinline[language=bash]{df_int}
    directory which contain the additional set of integrals with explicit Ewald correction;
    \item \lstinline[language=bash]!gf2! -- will generate the file \lstinline[language=bash]{df_ewald.h5} 
    that contains Ewald correction for integrals in second-order diagrams.
    \item \lstinline[language=bash]!gw! -- will generate the file \lstinline[language=bash]!AqQ.h5! 
    that contains a smooth interpolant for Coloumb integrals for the GW finite-size correction.
\end{itemize}
By default the \lstinline[language=bash]{ewald} correction is applied. 

\subsubsection{Convergence to a desired solution and convergence acceleration}

Self-consistent diagrammatic theories usually display multiple stationary points in their $\Phi-$ and $\Psi-$functional and very rarely have only only one global minimum. 
In certain cases, the existence of these multiple stationary points may be employed to our advantage since giving a starting solution that lies in the vicinity of the desired stationary point (local minimum) will predispose the algorithm to converge to this solution. This is usually done by searching for a desired solution in the mean-field (Hartree--Fock) method that produces multiple solutions, known as broken-symmetry solutions~\cite{Fukutome:UHF:81}. Subsequently, the weakly correlated perturbative fully self-consistent diagrammatic method often preserves the qualitative structure of the mean-field solution, thus, providing access to a broken-symmetry solutions of the Dyson equation. We have successfully used these correlated solutions for accurate evaluation of magnetic properties of molecules~\cite{Pokhilko:local_correlators:2021} and solids~\cite{Pokhilko:BS-GW:solids:2022} and even critical temperatures~\cite{Pokhilko:Neel_T:2022}.

To enable a smooth convergence to a desired solution and minimize the number of iterations required, its convergence path may be altered either by a constant 
iteration mixing, or by convergence acceleration methods such as variants of the direct inversion in the 
iterative subspace (DIIS)~\cite{PULAY1980393,Pulay1982,Pokhilko22}.

In iteration mixing, the self-energy is modified as:
\begin{align}
    \Sigma_{n} = \alpha \Sigma\left[G_{n-1}\right] + (1-\alpha)\Sigma_{n-1},
\end{align}
where $G_{n-1}$ and $\Sigma_{n-1}$ are Green's function and self-energy from the previous iteration and
$\alpha \in (0, 2)$ is an iteration mixing parameter (with $\alpha = 1$ meaning no mixing with previous iteration).

In DIIS, the self-energy at the given iteration is computed as a linear 
combination of the self-energies obtained in the $k$ previous iterations:
\begin{align}
    \Sigma_{n} = \sum_{i=0}^{k}c_{k} \Sigma_{n-1-k+i}.
\end{align}
Convergence acceleration methods often offer faster convergence and better stability.

To choose a convergence acceleration method, the user has to specify the parameter \lstinline[language=bash]{--mixing_type} which takes one of the four values \lstinline[language=bash]{NO_MIXING}, \lstinline[language=bash]{SIGMA_MIXING},
\lstinline[language=bash]{DIIS}, or \lstinline[language=bash]{CDIIS}.
Where CDIIS corresponds to commutator version of DIIS method~\cite{Pulay1982}.
If static iteration mixing is specified, the mixing weight $\alpha$ has to be provided by specifying \lstinline[language=bash]{--mixing_weight}.
If DIIS or CDIIS are chosen, the \Green/\WeakCoupling code uses static iteration mixing for the first few iterations, builds an extrapolation subspace, and continues with DIIS/CDIIS starting at the iteration specified by \lstinline[language=bash]{--diis_start}. The maximum number of subspace vectors used is specified by \lstinline[language=bash]{--diis_size}. Our implementation of DIIS and CDIIS is detailed in Ref.~\cite{Pokhilko22}

\section{Post processing and analysis}
The \Green/\WeakCoupling code provides post-processing tools such as analytical continuation to obtain spectral functions and thermodynamic utilities.

Analytical continuation using the Nevanlinna analytic continuation method \cite{Fei21} is performed by executing the program \lstinline[language=bash]{ac.exe} located at the installation path in the \lstinline[language=bash]{bin} subdirectory. The following parameters are needed: 

\begin{itemize}
    \item \lstinline[language=bash]{--grid_file} Sparse imaginary time/frequency grid file name
    \item \lstinline[language=bash]{--BETA} Inverse temperature
    \item \lstinline[language=bash]{--input_file} Name of the input file
    \item \lstinline[language=bash]{--output_file} Name of the output file
    \item \lstinline[language=bash]{--group} Name of the HDF5 group in the input file that contains imaginary time data. This group has to contain mesh information and data sets.
    \item \lstinline[language=bash]{--kind} Type of analytical continuation, currently only \lstinline[language=bash]{NEVANLINNA} is available.
\end{itemize}
The command \lstinline[language=bash]{ac.exe --help} will provide additional information.

\section{Example}\label{sec:examples}
We give step-by-step instructions on how to run \Green/\WeakCoupling for periodic silicon on a $6\times6\times6$ lattice. Results for a similar system are shown in Ref.~\cite{Yeh22}. Both \lstinline[language=bash]{mbpt.exe} and \lstinline[language=bash]{ac.exe} will be used.

Create a new simulation directory, and in it create the file \lstinline[language=bash]{a.dat} containing the following unit cell information (each line specifies one of the translation vectors in Angstroms): 
\begin{align*}
    \begin{matrix}
        0.0000,& 2.7155,& 2.7155 \\
        2.7155,& 0.0000,& 2.7155 \\
        2.7155,& 2.7155,& 0.0000
    \end{matrix}
\end{align*}

Then create a file \lstinline[language=bash]{atom.dat} containing atom positions in cell Angstroms unit cell:
\begin{align*}
\begin{matrix}
\text{Si}& 0.00000& 0.00000& 0.00000 \\
\text{Si}& 1.35775& 1.35775& 1.35775
\end{matrix}
\end{align*}

In a first step, obtain input parameters and an initial mean-field solution by running pySCF via the \lstinline[language=bash]{init_data_df.py} script:
\begin{verbatim}
python <source root>/green-mbpt/python/init_data_df.py         \
  --a a.dat --atom atom.dat --nk 6                             \
  --basis gth-dzvp-molopt-sr --pseudo gth-pbe --xc PBE         \
  --high_symmetry_path WGXWLG  --high_symmetry_path_points 300 \
  --job init sym_path 
\end{verbatim}

This will employ the \lstinline[language=bash]{gth-dzvp-molopt-sr} basis \cite{VandeVondele07} with the \lstinline[language=bash]{gth-pbe} pseudopotential, run a DFT calculation for the system  with a PBE exchange correlation potential and generate \Green/\WeakCoupling input. The high-symmetry path, which will be used in the analysis step, is set to \lstinline[language=bash]{WGXWLG}.

In a second step, we run the fully self-consistent GW approximation using the mbpt code. It is advantageous to run this step on an MPI cluster or on a GPU:
\begin{verbatim}
mpirun -n <number of cores> <install dir>/bin/mbpt.exe --scf_type=GW --BETA 100       \
  --grid_file ir/1e4.h5 --itermax 20 --results_file Si.h5 \
  --mixing_type CDIIS --diis_size 5 --diis_start 2 \
  --high_symmetry_output_file Si_hs.h5 --jobs SC,WINTER
\end{verbatim}

This sets the inverse temperature to $\beta=100$ Ha$^{-1}$, employs an Intermediate Representation grid~\cite{PhysRevB.96.035147,PhysRevB.101.035144} with grid parameter $ \Lambda = 10^4$, and runs for 20 iterations using DIIS convergence acceleration.

\begin{figure}[tbh]
    \centering
    \includegraphics[width=.9\textwidth]{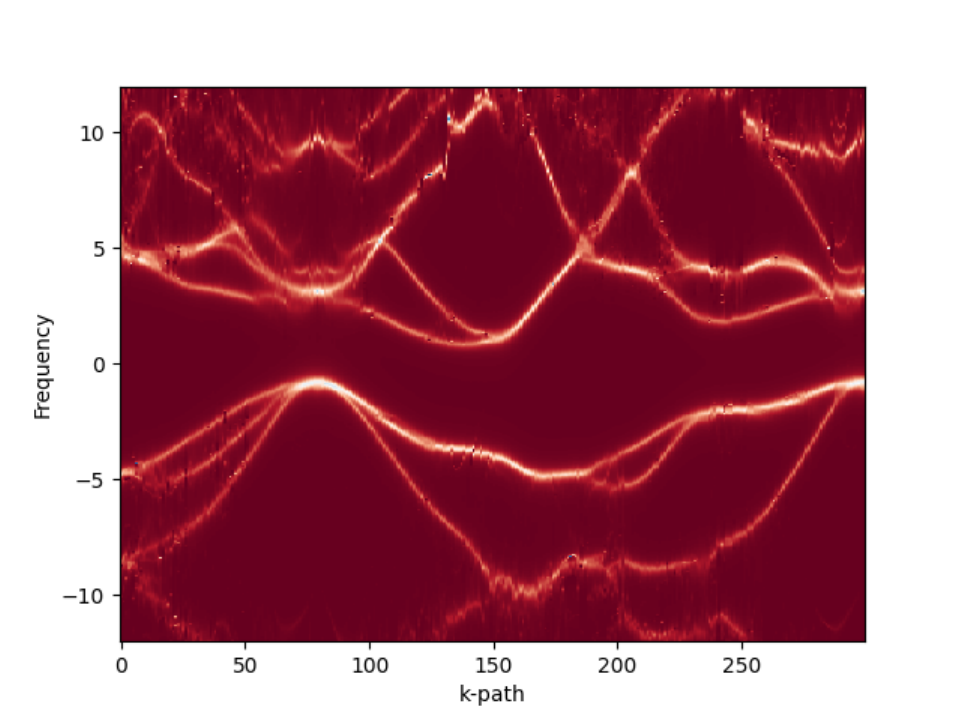}
    \caption{Band structure for silicon obtained from GW calculation on $6\times6\times6$ lattice and extrapolated along high-symmetry path.}
    \label{fig:bands}
\end{figure}

The simulation results are stored into the file \lstinline[language=bash]{Si.h5}, and the band-path interpolation along the high-symmetry path in \lstinline[language=bash]{Si_hs.h5}. 

In a third step, we use the post-processing analytic continuation to obtain the spectral function:
\begin{verbatim}
<install dir>/bin/ac.exe  --BETA 100 --grid_file ir/1e4.h5    \
   --input_file Si_hs.h5 --output_file ac.h5 --group G_tau_hs \
   --e_min -10.0 --e_max 10.0 --n_omega 10000 --eta 0.01
   --kind Nevanlinna
\end{verbatim}
This will run analytical continuation for all momentum points along the high-symmetry path. The output will be stored in the group
\lstinline[language=bash]{G_tau_hs} of the output HDF5 file \lstinline[language=bash]{ac.h5}.

Finally, a plot for the band structure can be obtained with the \lstinline[language=bash]{plot_bands.py} script:
\begin{verbatim}
python <source root>/green-mbpt/python/plot_bands.py \
        --input_file ac.h5 --output_dir bands
\end{verbatim}
This will read the analytically continued data and plot it to an \lstinline[language=bash]{<output_dir>/bands.png} file. In addition, it will create a plain-text data file containing the spectal function for every k-point along the chosen path inside the \lstinline[language=bash]{<output_dir>} directory.
The resulting band structure plot is shown in Fig.~\ref{fig:bands}.

Other examples, such as calculation of ionization potentials in molecular systems~\cite{doi:10.1021/acs.jctc.3c01279} or using spin-orbit coupling~\cite{doi:10.1021/acs.jctc.4c00075}, please check the \Green/\WeakCoupling web-site.

\section{Conclusion}\label{sec:summary}
We have presented an implementation of self-consistent finite-temperature many-body perturbation theory for molecules and solids, along with analytic continuation tools.
The software provided as part of the \Green project provides a way to perform diagrammatic simulations of solids and molecules without adjustable parameters.
We have provided scripts and examples for software installation, input preparation, solution of the many-body problem, and analytic continuation analysis.
Our implementation is licensed with the permissive MIT license that allows users to use, copy, modify, merge, publish, distribute, sublicense, and sell copies of software.

\section{Acknowledgments}
This material is based upon work supported by the National Science Foundation under Grant No.~OAC-2310582.

\bibliographystyle{elsarticle-num.bst}
\bibliography{main}

\end{document}